\documentclass[preprint2,twoside]{hwo}

\usepackage{graphicx}

\newcommand{\muas}{{\mu{\rm as}}}

\input{hwo.h}

\begin{document}

\title{\textbf{\LARGE Masses of Potentially Habitable Planets Characterized by the Habitable Worlds Observatory}}

\author {\textbf{\large Kaz Gary$^1$, B. Scott Gaudi$^{1}$,  Eduardo Bendek$^2$}}
\affil{$^1$\small\it Department of Astronomy $\vert$ The Ohio State University, Columbus, OH 43210, USA}
\affil{$^2$\small\it NASA Ames Research Center, Moffett Field, Mountain View, CA 94035-1000, USA}

\author{\small{\bf Contributing Authors:} Ty Robinson (Lunar $\&$ Planetary Laboratory $\vert$ University of Arizona, Tucson, AZ 85721, USA $\&$ Habitability, Atmospheres, and Biosignatures Laboratory $\vert$ University of Arizona, Tucson, AZ 85721, USA), Renyu Hu (Jet Propulsion Laboratory $\vert$ California Institute of Technology, Pasadena, CA 91011, USA $\&$ Division of Geological and Planetary Sciences $\vert$ California Institute of Technology, Pasadena, CA 91125, USA), Breann Sitarski (NASA Goddard Space Flight Center, 8800 Greenbelt Rd, Greenbelt, MD 20771, USA), Aki Roberge (NASA Goddard Space Flight Center, 8800 Greenbelt Rd, Greenbelt, MD 20771, USA) and Eric Mamajek (Jet Propulsion Laboratory $\vert$ California Institute of Technology, Pasadena, CA 91011, USA)
}




\begin{abstract}

    Constraints on the masses of exoplanets directly imaged and characterized by the Habitable Worlds Observatory (HWO) are crucial for categorizing these planets and interpreting their spectra. In particular, achieving a mass measurement with a precision of approximately 10$\%$ or better may be necessary to identify the dominant gaseous species in the atmospheres of Earth-like planets. This, in turn, is essential for assessing their potential habitability and interpret potential biosignatures \citep{Damiano:25}. Space-based astrometry will be required to measure the masses of planets in face-on systems, or planets orbiting hot and rapidly rotating or highly active stars. Astrometric uncertainties are dominated by the number and magnitude of background reference stars needed to precisely measure the astrometric wobble of the target star induced by the planet. To that end, we propose a program to measure the masses of Earth analogs orbiting HWO target stars with ultra-high-precision astrometry obtained with the HWO high-resolution instrument. We assess the photon-noise error budget for these observations.  We find that, for a field of view spanning a few square arcminutes, the astrometric uncertainty due to the number and brightness of reference stars dominates the photon-noise error budget, particularly for targets near the Galactic poles. We explored the impact of filter choice and location in the sky on the photon-noise astrometric uncertainties by simulating the magnitude distribution of reference stars across different filters at a range of galactic longitudes and latitudes. We find that a $\sim 200$-day survey in the Gaia G band consisting of 100 epochs per target star of distributed over the 5-year prime mission with a 6m aperture HWO equipped with a $6'\times 6'$ field-of-view would be required to achieve the photon-noise sensitivity to measure the masses of the $\sim 40$ Earth-mass habitable-zone planets to $\sim 10\%$.
    \\
    \\
\end{abstract}

\vspace{2cm}

\section{Science Goal}

{\bf Can we identify habitable conditions from the spectra of nearby Earthlike planets?}

The Habitable Worlds Observatory (HWO) will search for habitable conditions and biosignatures on planets orbiting nearby stars by obtaining UV/Optical/near-IR spectra of the planets \citep{Feinberg:2024}.  However, these spectra cannot be properly interpreted in isolation.  Rather, the full context of the properties of the planet, its orbit, and the host planetary system will be needed to robustly assess the potential for habitability and life.  

In particular, knowledge of the mass of a directly-imaged planet is essential.  The mass of a planet is required to broadly categorize the nature of the planet as terrestrial, super-Earth, Neptune, or gas giant. For example, for transiting planets, the transition from rocky to gaseous worlds appears to occur around a few Earth masses \citep{Rogers:2015,Fulton:2017}.  For a terrestrial planet, mass helps dictate the origin and evolution of the atmosphere and influences atmospheric dynamics, outgassing, and escape.  Indeed, the "cosmic shoreline" that divides planets with and without atmospheres depends on instellation and escape velocity \citep{Zahnle:2017}, the latter of which depends on mass. 

\begin{figure*}[ht!]
    \centering
    \includegraphics[width=0.75\textwidth]{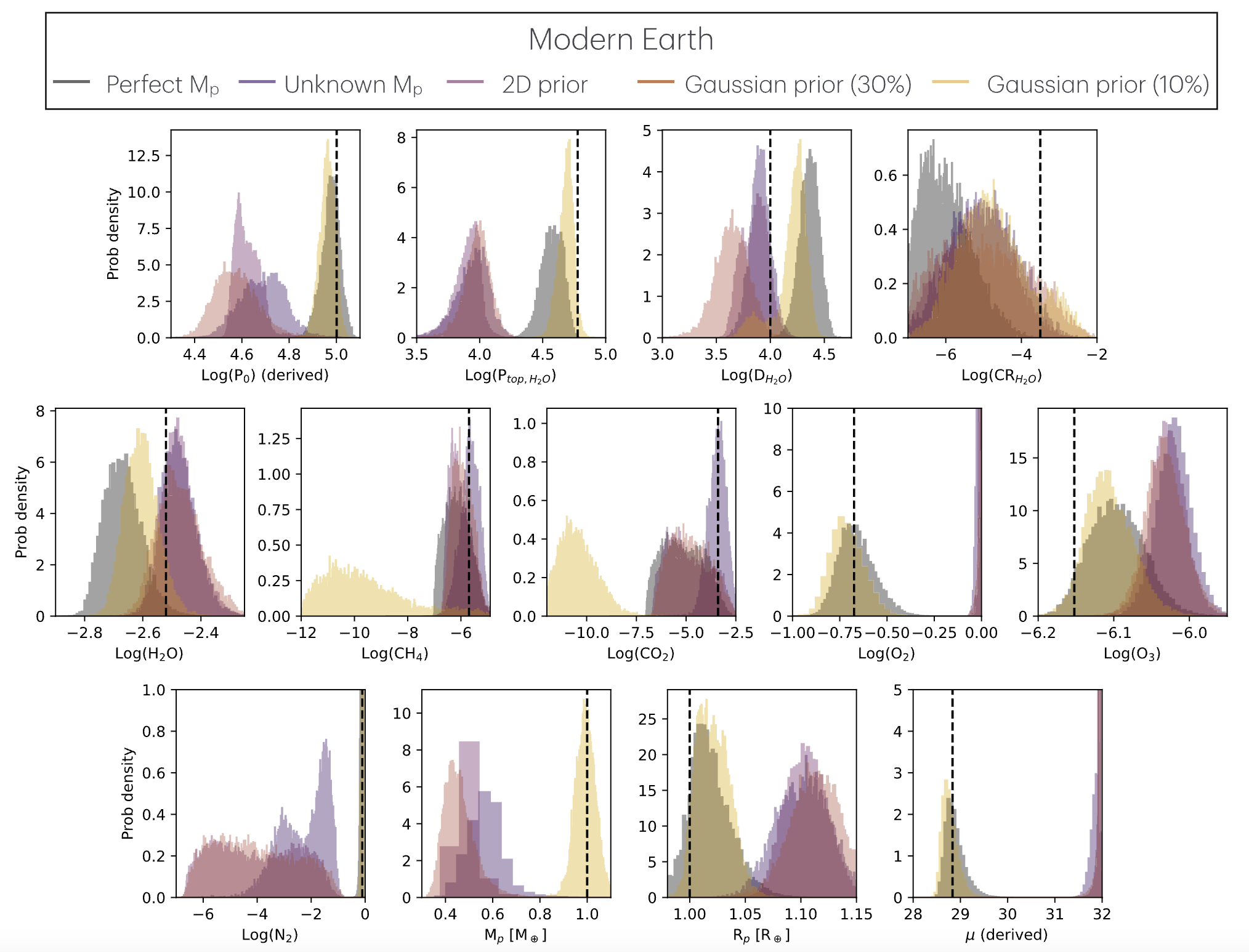}
    \caption{\small {\bf A prior on the mass of a directly-imaged Earthlike planet is needed to accurately interpret the reflected-light spectrum.} Each panel shows the posterior probability disribution for the parameters used in the retrieval of a simulated reflected-light spectrum of a modern-day Earthlike planet. A prior on the planetary mass of 10\% is needed to correctly identify the dominant background atmospheric gas.
        Figure from \cite{Damiano:25}, used with permission.}
    \label{fig:mass_prior}
\end{figure*}

Furthermore, an estimate of the mass of a directly imaged planet helps to most accurately interpret its reflected light spectra \citep{Nayak:2017,Damiano:25}, and in particular to most accurately determine the abundances of the atmospheric constituents of potentially habitable planets and thus robustly determine if they indeed host habitable conditions.  For example, \citet{Damiano:25} show that a prior on the planetary mass of 10\% is needed to correctly identify the dominant background atmospheric gas and thus properly characterize the habitability of modern-day and Archean Earth-like planets (see \citealt{Damiano:25} and Figure \ref{fig:mass_prior}).  


It is likely that planet mass will be one of the critical physical parameters that will not be routinely available for most of the potentially habitable planets that HWO will characterize, and thus a robust strategy for measuring these masses needs to be developed.  If we set the minimum requirements for the habitability of an Earthlike planet to be the ability to host liquid water on its surface and a specific set of atmospheric constituents, then the `first-order' parameters that must be constrained to assess habitability are the surface temperature, surface pressure, and atmospheric abundances.  

Retrievals on simulated \citep{Young:2024} and real \citep{Robinson:2023} reflected light observations of Earth-like worlds have revealed that temperature broadening of gas absorption bands yields constraints on the thermal environment of the atmosphere, with typical uncertainties spanning 10—100 K. A surface temperature can be further constrained via modeling, using the host spectral energy distribution, the orbit of the planet, and constraints from reflected-light observations on the atmospheric composition.  The luminosity and spectral energy distributions of the HWO host stars are already relatively well-measured (with the possible exception of their high-energy emission, which is of lesser importance for models of planetary climate).  The orbit of the planets will generally be measured by HWO for promising targets. Greenhouse gas opacity can be inferred with a constraint on the abundances from HWO high-contrast imaging and spectroscopy, although constraints on some relevant greenhouse gases may only be an upper limit as gas absorption bands at optical/near-infrared wavelengths are generally weaker than bands at thermal infrared wavelengths. The ability to reverse-model the climate of a planet with incorporation of HWO-like observational uncertainties on gas abundances has yet to be demonstrated. 

The surface pressure and atmospheric abundances, and atmospheric mean molar weight (all critical to climate modeling as well as environmental interpretation) can be inferred from the reflected-light spectrum as follows.  The strength of the absorption features due to species $i$ depends on the optical depth $\tau_i$, which is roughly $\tau_i \sim \kappa_i M_{c,i}\sim \sigma_i m_i N_{c,i}$, where $\kappa_i$, $\sigma_i$ $m_i$, $N_{c,i}$, $M_{c,i}$, are the opacity (per gram), absorption cross section (per molecule), mass, column number density, and column mass density of species $i$.  From hydrostatic equilibrium, we have the following.
\begin{equation}
    M_{c,i} \sim \frac{P_i}{g},
\end{equation}
where $P_i$ is the partial pressure of the species $i$ and $g=GM_p/R_p^2$ is the surface gravity. Thus, the unknowns are the individual gas partial pressures, the planetary radius, the planetary mass, and the total atmospheric pressure (which can be larger than the sum of the partial pressures of the absorbing gases due to the presence of a background gas). With absorption features measured for each active gas, there are still three unknowns: planetary radius, planetary mass, and total pressure. 

Thus, to infer the partial pressure of each species from the absorption features, as well as the mean molecular weight of the atmosphere, one must have a constraint on $g$. The radius of the planet can be inferred from its flux $F_p$ relative to the star $F_s$, its geometric albedo $A_g$, phase function $\Phi(\alpha)$, and semimajor axis $a$ through
\begin{equation}
\frac{F_p}{F_*}= A_g \Phi(\alpha) \left(\frac{R_p}{a}\right)^2,
\end{equation}
where $\alpha$ is the phase angle.  The fluxes $F_p$ and $F_s$ are directly measured quantities, and $a$ and $\alpha$ are known from the orbit of the planet.  The geometric albedo $A_g$ and phase function are known in the case of Rayleigh scattering\footnote{Although we note that atmospheric hazes and clouds may complicate the interpretation of the spectra.}, thus yielding a constraint on radius \citep{Feng:2018}. This leaves a correlation between total atmospheric pressure and planetary mass\footnote{Fpr rocky planets. the mass-radius relationship depends weakly on composition, and thus surface gravity is tightly correlated with mass.} \citep{Salvador:24,Damiano:25}. As the total column number density (i.e., summed over all species) is equal to $P/m/g$, the atmospheric mean molar weight ($m$) also cannot be well constrained.  

Thus, measuring planetary mass can unlock information about key atmospheric bulk parameters, especially pressure and mean molar weight. However, mass is the only parameter that is not constrained from existing knowledge of the system or will not be constrained from the HWO coronographic observations themselves, at least in the case of Earth-like planets whose reflectance spectra are dominated by Rayleigh scattering. 

Returning to the issue of the degeneracy in the identification of the dominant background gas identified by \citet{Damiano:25} when one does not have a tight constraint on the planet's mass, it is worth noting that how problematic one finds the existence of this degeneracy depends on one's prior assumption for the range of plausible background gases. For example, if one restricts oneself to a relatively short list of gases that can plausibly make the bulk atmosphere constituents on rocky planets, e.g., H$_2$, He, H$_2$O, N2, CO, O$_2$, Ar, CO$_2$, SO$_2$, then if $\mu$ is constrained to be $\sim 28$ from the spectrum, one just needs to distinguish between N$_2$ and CO.  In this case, sufficient NIR coverage of the reflectance spectra should be able to distinguish between these two scenarios \citep{Hall:2023}. Of course, one could imagine a scenario in which a combination of two inert gases - say He and Ar - combine to provide $\mu \sim 28$, but this scenario may be a priori implausible.

\section{Science Objective}

\textbf{Measuring the masses of potentially habitable planets.}

\begin{figure*}[ht!]
    \centering
       \includegraphics[width=0.44\linewidth]{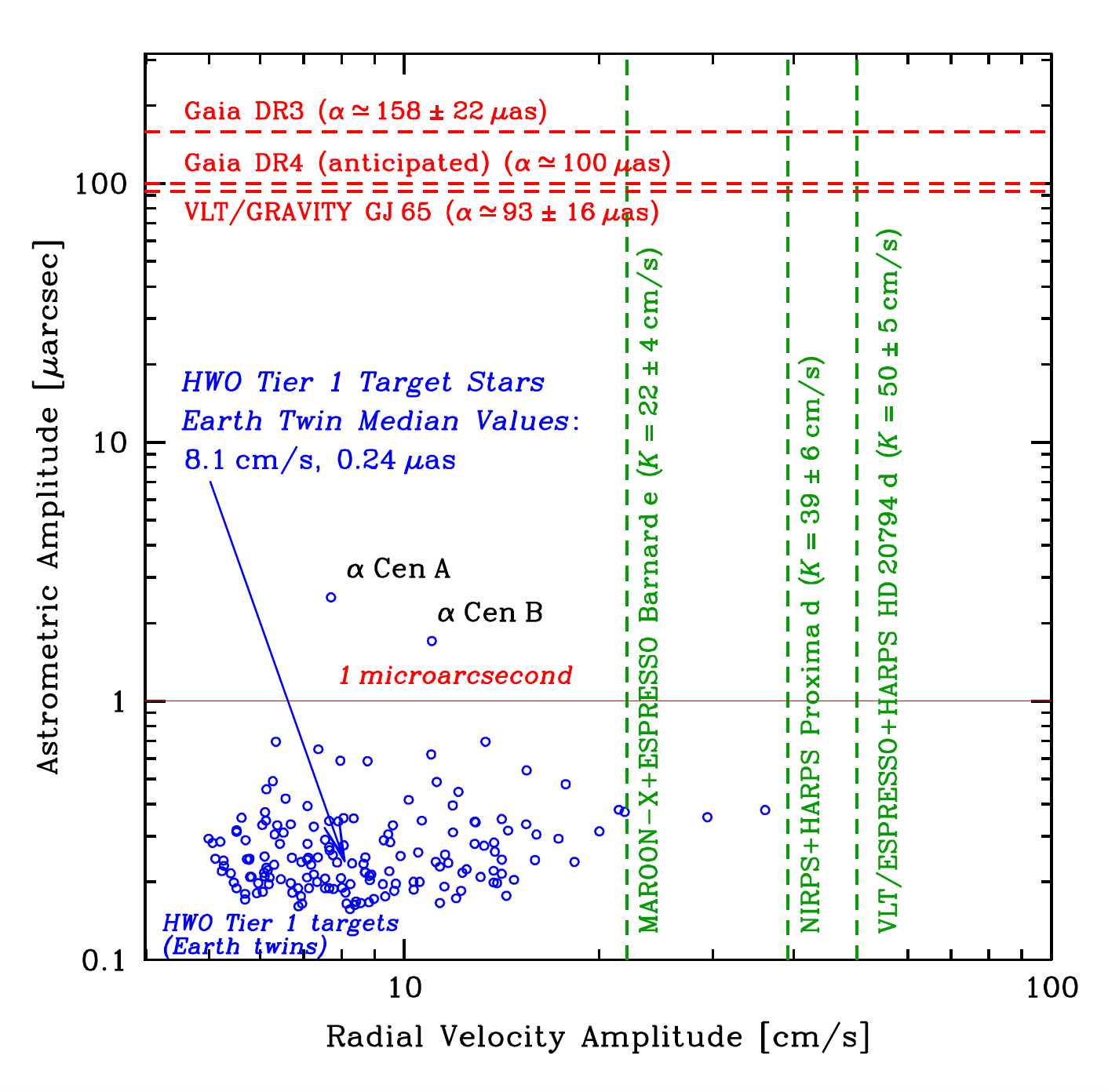}
        \includegraphics[width=0.55\linewidth]{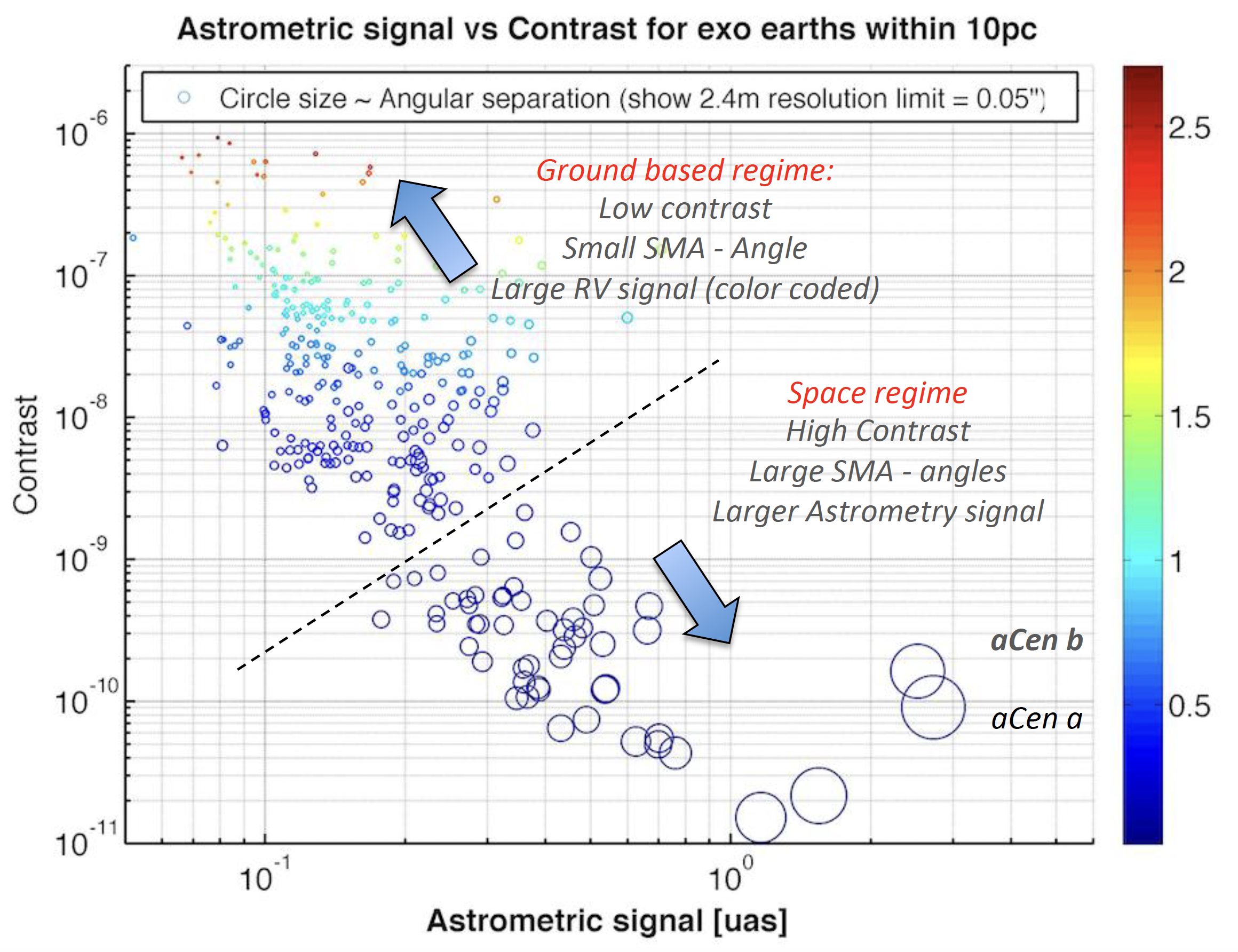}
        \caption{\small {\bf The astrometric and RV signals of Earth analogs orbiting nearby stars are small and below the current state of the art, while the astrometric signals of such planets are positively correlated with their angular separation but anti- correlated with their contrast and RV signal.} (left) The amplitude of the astrometric reflex signal versus the radial velocity reflex signal for Earth-mass planets receiving an Earth-like instellation orbiting a sample of nearby stars that are likely to be targets for HWO direct imaging. Credit: Eric Mamajek, used with permission. (right) Contrast versus astrometric signal for a similar sample of stars hosting Earth-like planets.  The size of the circles are proportional to the angular separation of the planet and the colors of the circles are proportional to the RV amplitude. 
 }\label{fig:astrovesrv}
\end{figure*}

The only way of routinely measuring the masses of planets orbiting the nearby bright FGK and early M stars that are the targets for HWO is through measurement of the reflex motion of the star due to the planet's orbit via precise radial velocities (RVs) or astrometry \citep{Bendek:2018}. The signals resulting from an Earth analog orbiting a sunlike star are extremely small. Consider a planet with a mass of $M_p$ orbiting a star with a mass of $M_*$ on a circular orbit with semimajor axis $a$ and inclination $i$.  Assume that the system is at a distance of $d$.  Then, assuming $M_p\ll M_*$, the RV amplitude of the reflect motion due to the planet is
\begin{equation}
K_* = 8.9~{\rm cm/s}
\left(\frac{M_p \sin{i}}{M_\oplus}\right)
\left(\frac{M_*}{M_\odot}\right)^{-1/2}
\left(\frac{a}{{\rm au}}\right)^{-1/2} ,
\end{equation}
whereas the astrometric signal due to the planet is
\begin{eqnarray}
\begin{split}
        \alpha_{*,\perp} &= 0.3~\mu{\rm as}
        \left(\frac{M_p}{M_\oplus}\right)
        \left(\frac{M_*}{M_\odot}\right)^{-1}
        \left(\frac{a}{{\rm au}}\right)
        \left(\frac{d}{{\rm pc}}\right)^{-1},
        \\ 
        \alpha_{*,\parallel}&=\alpha_{*,\perp} \cos{i},
        \end{split}
\end{eqnarray}
where $\alpha_{*,\perp}$ and $\alpha_{*,\parallel}$ are the components of the astrometric reflex motion perpendicular and parallel to the projected orbit normal.  Figure \ref{fig:astrovesrv} shows the magnitude of these signals for a sample of nearby stars that are the likely targets for HWO assuming a planet of $M_p=M_\odot$ receiving an equivalent instellation as Earth. For both RV and astrometry, these signals are generally well below the current state-of-the-art (see Figure \ref{fig:astrovesrv} and Table~\ref{tab:mass_parameters}).

\section{Physical Parameters}

If we assume that there are $N_d$ uncorrelated, Gaussian-distributed measurements of the RV of the star with precision $\sigma_{\rm RV}$ that are densely and uniformly spaced in the phase of the planet's orbit, then the signal-to-noise ratio (SNR) with which the RV signal is measured is,

\begin{equation}
    {\rm SNR}_{\rm RV} \simeq \sqrt{\frac{N_d}{2}}\frac{K_*}{\sigma_{RV}}.
\end{equation}

Assuming that $M_*$ is precisely known, the uncertainty in $M_p\sin{i}$ is approximately
\begin{equation}
\frac{\sigma_{M_p\sin{i}}}{M_p\sin{i}}
\sim 
\frac{\sigma_{K_*}}{K_*}
\sim ({\rm SNR}_{\rm RV})^{-1}.
\end{equation}

Therefore, in order to measure the mass of an Earth-mass planet orbiting a sunlike star to $\sim 10\%$, one needs, e.g., $N_d=200$ RV measurements, each with a measurement precision of $\sigma_{\rm RV}\sim 10~{\rm cm/s}$.  Other sources of uncertainty and error must ultimately be controlled or removed at the level of $\sim K_*/10 \sim 1~{\rm cm/s}$ to measure the mass of the planet to $\sim 10\%$.

Similarly, if we assume that there are $N_d$ two-dimensional, uncorrelated, Gaussian-distributed measurements of the relative astrometric position of the star with precision $\sigma_{\rm ast}$ in each dimension that are densely and uniformly spaced in the phase of the planet's orbit, then the signal-to-noise ratio (SNR) with which the astrometric signal is measured is
\begin{equation}
    {\rm SNR}_{\rm ast} \simeq \sqrt{{N_d}\frac{1+\cos^2{i}}{2}}\frac{\alpha_*}{\sigma_{\rm ast}}.
\end{equation}
Assuming that $M_*$ and $d$ are precisely known, the uncertainty in $M_p\sin{i}$ is approximately
\begin{equation}
\frac{\sigma_{M_p}}{M_p}
\sim 
\frac{\sigma_{\alpha_*}}{\alpha_*}
\sim ({\rm SNR}_{\rm ast})^{-1}.
\end{equation}

Therefore, in order to measure the mass of an Earth-mass planet in a face-on orbit around a sunlike star to $\sim 10\%$, one needs, e.g., $N_d=100$ astrometric measurements, each with a measurement precision of $\sigma_{\rm ast}\sim 0.3~\muas$.  Other sources of uncertainty and error must ultimately be controlled or removed at the level of $\sim \alpha_*/10 \sim 0.03~\muas$ to measure the mass of the planet to $\sim 10\%$.


\begin{table*}[ht!]
    \centering
    \caption[Performance Goals]{\small \textbf{Current capabilities and future milestones for mass measurements of habitable zone planets orbiting nearby FGK stars}. Masses are derived from RV and astrometry observations listed in Table 2. We assume $a=1 {\rm AU}$, $d=10~{\rm pc}$, $M=M_\oplus$, and circular orbits.}
    \label{tab:mass_parameters}
    \begin{tabular}{ccccc}
   \hline
        State of the Art  & Incremental Progress & Substantial Progress & Major Progress \\
 & (Enhancing) & (Enabling) & (Breakthrough) \\
       \hline
           60~$M_\oplus$ (RV) & 10~$M_\oplus$ & 1~$M_\oplus$ & 0.1~$M_\oplus$ \\
         20~$M_\oplus$ (Ast.) & 10~$M_\oplus$ & 1~$M_\oplus$ & 0.1~$M_\oplus$ \\
    \end{tabular}
    \caption{\small \textbf{Current capabilities and future milestones for RV and astrometry measurements of habitable zone planets orbiting nearby FGK stars}. For RV, the mass limit corresponds to $M_p \sin{i}$. For astrometry, the mass limit assumes face-on orbits. We assume a state of the art for the minimum detectable RV and astrometric signals to be $K_{*,{\rm min}}=6~{\rm m/s}$ \citep{harada:2024} and $\alpha_{*,{\rm min}}=30~\mu as$ \citep{Eyer2024}. We define incremental progress as the ability to distinguish between rocky and gas-rich planets (e.g., \citealt{Rogers:2015}), but not be able to measure the mass, enabling progress as the ability to measure the mass of an Earth twin, and breakthrough progress as the ability to measure the minimum mass of a planet that is likely to be habitable for many Gyrs, which we define to be a Mars-mass planet ($M_p\simeq 0.1~M_\oplus$).}
    \label{tab:physical_parameters}
    \begin{tabular}{ccccc}
   \hline
        State of the Art  & Incremental Progress & Substantial Progress & Major Progress \\
 & (Enhancing) & (Enabling) & (Breakthrough) \\
       \hline
           RV: 6~{\rm m/s} & 1~{\rm m/s} & 10~{\rm m/s} & 0.1 ~{\rm m/s} \\
         Ast: $30~\muas$ & $3~\muas$ & $0.3~\muas$ & $0.03~\muas$ \\
    \end{tabular}
\end{table*}

Note that a measurement of the planet mass via both astrometry and radial velocity requires a measurement or a constraint on the mass of the host star. In this case of astrometry, it also requires a measurement of the distance to or parallax of the system. In general, we can write
\begin{equation}
    \left(\frac{\sigma_{M_p}}{M_p}\right)^2 = \left(\frac{\sigma_{K_*}}{K_*}\right)^2+\left(\frac{\sigma_{M_*}}{M_*}\right)^2\,\, {\rm (RV)},
\end{equation}
and
\begin{equation}
    \left(\frac{\sigma_{M_p}}{M_p}\right)^2 = 
    \left(\frac{\sigma_{\alpha_*}}{\alpha_*}\right)^2
    +\left(\frac{\sigma_{M_*}}{M_*}\right)^2
    +\left(\frac{\sigma_{d}}{d}\right)^2\,\, {\rm (Ast.)}.
\end{equation}
The distances to HWO target stars are already known quite precisely, with an average uncertainty of $\sim 0.1\%$.  However, the masses of the HWO target stars are much less precisely known.  In general, measuring or inferring the masses of isolated field stars is very difficult. It is worth noting that it might be possible to measure the masses of the HWO target stars to $\sim 0.1\%$ by measuring the gravitational deflection of background stars due to gravity of the foreground star using HWO astrometry \citep{Wright:2023}.

\subsection{The Complementarity of Radial Velocity and Astrometry and the Need for Both}

The accuracy of RV measurements is limited by photon noise, stellar activity, and instrumental effects. 
Achieving the $\sim 1~{\rm cm/s}$ accuracy required to detect the RV signal of an Earth analog is extremely challenging. Instrumental stability and photon noise aside, stellar activity begins to dominate RV variations below a level of $\sim 1~{\rm m/s}$, even for relatively quiet Sunlike stars \citep{Fischer:2016,Crass:2021}. Strategies for removing the stellar activity and controlling the instrumental contribution to the $\sim 10~{\rm cm/s}$ level or better have been developed and are being implemented (see \citealt{Fischer:2016,Crass:2021} and references to these works). The state of the art is currently at least an order of magnitude away from the required accuracy (see Fig.~\ref{fig:astrovesrv} and, e.g., \citealt{Figueira:2025}).  Nevertheless, progress is being made, and RV remains one of the most promising methods for measuring masses of terrestrial planets in the habitable zones of nearby stars, especially for relatively low-mass hosts.  It is partially for this reason that the NASEM 2018 Exoplanet Science Strategy report \citep{NAP25187} prioritized a "strategic initiative in extremely precise radial velocities to develop methods and facilities to measure the masses of temperate terrestrial planets orbiting Sun-like stars."

  \begin{figure*}[ht!]
        \centering
       \includegraphics[width=0.48\linewidth]{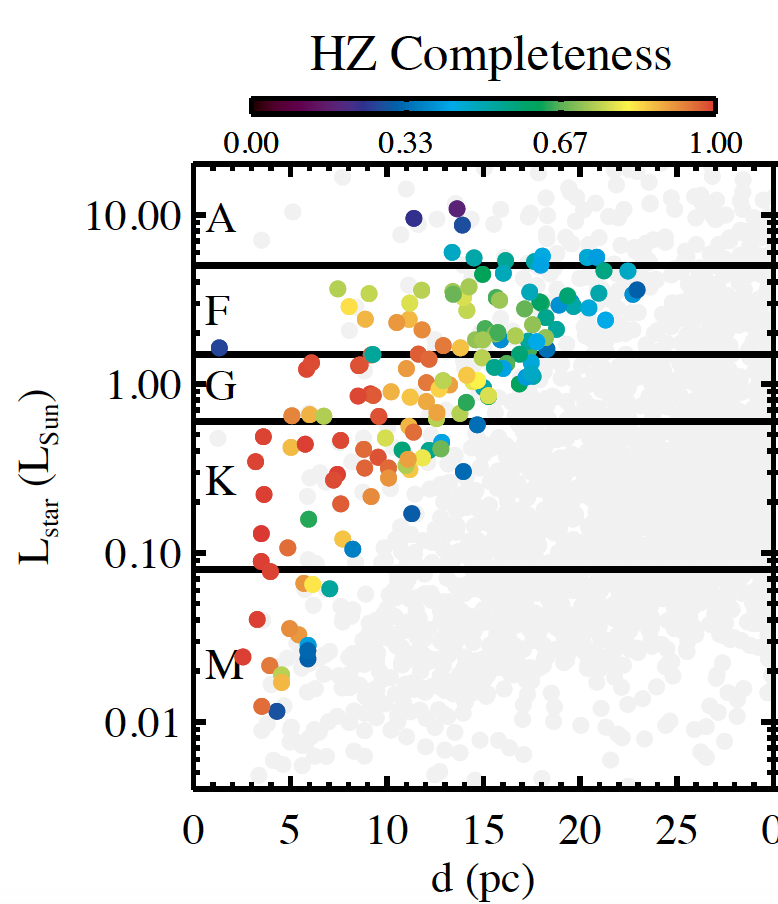}
        \includegraphics[width=0.51\linewidth]{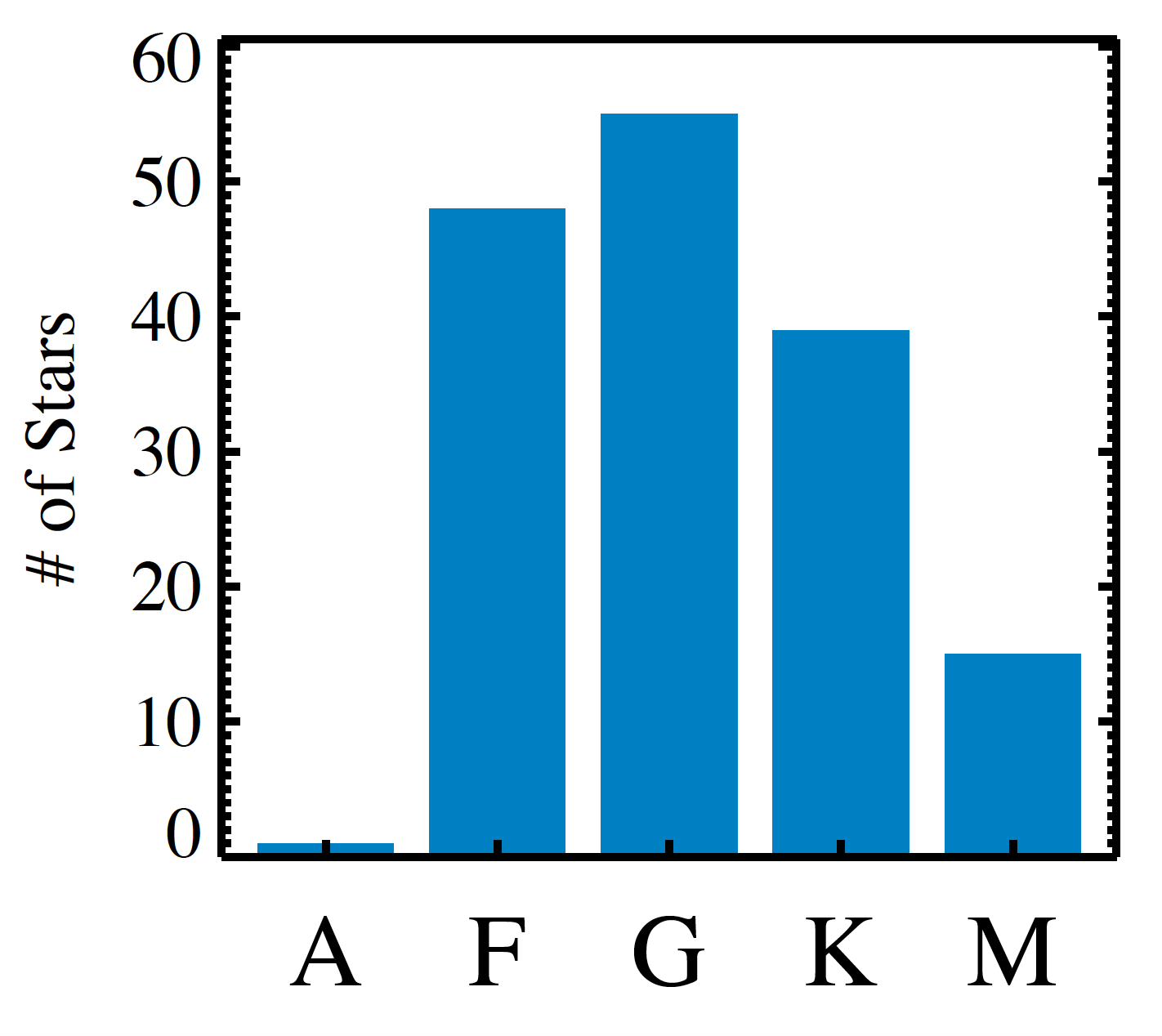}
        \caption{\small {\bf Roughly 30\% of the LUVOIR-B (and thus HWO) target sample will be of spectral type A or F, and thus be too hot and/or rapidly rotating for very high precision RV measurements.} (left) The colored points show the stellar luminosity versus distance for the 158 nearby stars the LUVOIR-B target sample, which will be similar to the HWO sample. The points color coded by habitable zone (HZ) completeness (right) Distribution of spectral types for the LUVOIR-B sample. Figure from \cite{LUVOIR}, used with permission.
 }\label{fig:LOVOIRB}
    \end{figure*}

However, measuring sub-m/s RV signals will remain extremely difficult or impossible for a subset of stars.  These include relatively active stars, but also stars for which the RV precision is ultimately
limited by the photon noise, which depends on signal-to-noise ratio of the spectra used to measure the RV and the number and depth of the resolved spectral lines. For hot stars with effective temperatures above the Kraft break of $T_{\rm eff} \gtrsim 6300~{\rm K}$  \citep{Kraft:1967,Beyer:2024}, corresponding to spectral types earlier than roughly F7 \citep{Pecaut:2013}, there is a paucity of spectral lines because of their hot photospheres and thus a lack of RV information \citep{Beatty:2015}. Furthermore, these stars have thin convective envelope and thus relatively weak (ordered) magnetic fields, and as a result do not spin down due to magnetic breaking and tend to be rapidly rotating ($v\sin{i} \gtrsim  20~{\rm km/s}$).  

The net result is that A and early F stars have fewer, weaker, and broader spectral lines, and thus will never be amenable to very high-precision RV measurements \citep{Beatty:2015} at the level need to detect Earth analogs. Although the HZs of these stars generally have larger angular separations (at fixed distance) than for GKM stars, the contrasts for Earth-like HZ planets are smaller (Fig.~\ref{fig:astrovesrv}). Nevertheless, $\sim 30\%$ of the LUVOIR-B target sample are A and F stars \citep{LUVOIR}, and 20\% of the Tier 1 targets from the HWO Target Stars \& Systems Working Group list are earlier than F7 \citep{Tuchow2025}.  Thus, astrometry will be needed to measure the masses of the potentially habitable planets orbiting these hot stars (as well as active stars).

It is worth noting that astrometry is complementary to RV in the astrometric signal for planets in the HZ increases with host star mass, in contrast to the RV signal.
\begin{eqnarray}
    \begin{split}
        K_{\rm HZ} &\sim 9~{\rm cm/s} 
      \left(\frac{M_p}{M_\oplus}\right)
      \left(\frac{M_*}{M_\odot}\right)^{-1.675},\\
        \alpha_{\rm HZ} &\sim 0.3~\mu{\rm as}  
     \left(\frac{M_p}{M_\oplus}\right)
     \left(\frac{M_*}{M_\odot}\right)^{1.35}
     \left(\frac{d}{10~{\rm pc}}\right)^{-1},
    \end{split}
\end{eqnarray}
where we have assumed that the bolometric luminosity of the star scales as $L_*/L_\odot = (M_*/M_\odot)^{4.7}$ and $a_{\rm HZ} \propto L_*^{1/2}$.  Astrometry is also complementary to RV in that the SNR of astrometric measurements is largest for face-on orbits, whereas in the case of RV it is largest for edge-on orbits.  Ultimately, it would be advantageous to use both RV and astrometry to improve the signal-to-noise ratio and control systematics, where possible. 

Achieving the $\sim 0.03~\mu{\rm as}$ accuracy required to detect the astrometric signal of an Earth analog is also extremely challenging.  However, for astrometry, intrinsic astrometric jitter due to stellar activity (e.g. starspots) is thought to not dominate the astrometric error budget \citep{Makarov:2009, Meunier:2022}.  Rather, the uncertainties due to photon noise and systematic errors that arise from time-varying optical distortions and inhomogeneous detector response are likely to dominate the precision and accuracy of the astrometric measurements.  

Ultimately, even if all sources of astrophysical astrometric variability and systematic errors can be eliminated, the photon noise provides the fundamental limit to the precision with which the masses of planets can be measured astrometrically.  Therefore, the first step in constructing the error budget for astrometric measurements of the masses of planets with HWO is to assess the uncertainty due to photon noise. Here we define ``photon noise" broadly.  As HWO will obtain time-series astrometry of the target star {\it relative} to a set of background stars using images obtained with a high-resolution imaging camera (HRI), the contributions to photon noise include the ability to centroid the point spread functions (PSFs) of the target and reference stars, and the number and brightness of the reference stars.  In addition, the target and reference stars will also have astrometric motion due to proper motion and parallax, and the reference stars may also have unseen companions. Finally, the reference stars will be deflected because of the gravity of the foreground star. All of these effects must be modeled self-consistently. The resulting mass uncertainty will be affected by covariances of the astrometric signal of interest with these 'nuisance' signals, as well as potentially any inaccuracies and/or degeneracies in the modeling of these time series of astrometric measurements. 

HWO is a promising observatory for performing ultra-high-precision astrometry due to the exquisite telescope stability imposed by the requirement to obtain very high contrast ($\sim 10^{-10}$) coronography.  However, it is not even clear whether astrometric measurements at the photon noise limit allow for the determination of the masses of planets at the requisite level of accuracy.  In particular, the nominal design for HRI has a relatively small FOV ($\sim 6' \times 6'$), and thus the number of reference stars will be limited.  Furthermore, the nominal HWO design is not ideally suited for high-precision astrometric measurements, and the control of systematic errors due to time-varying optical distortion and changes in detector response may require hardware modifications \citep{Guyon:2012,Bendek:2013a,Bendek:2013b,Bendek:2020,Crouzier:2016}.

In this SCDD, we estimate the magnitude of the photon-noise statistical astrometric uncertainty for HWO and use this to define the requirements for the mission architecture (for example, aperture size, field of view), astrometric survey strategy (for example, total observation time, filter choice, number of epochs, cadence), and systematic error control necessary to achieve 10\% mass measurements for Earth-like planets.

\section{Description of Observations}

HWO will measure the difference in position of the target star relative to the background star by performing astrometry on images of the target field taken by an imaging camera.  The precision with which the centroid of an (approximately Gaussian) point spread function (PSF)  of a star can be measured is roughly
\begin{equation}
    \sigma_{\rm ast} \simeq \frac{\rm FWHM}{2\sqrt{2}Q_\gamma}
\end{equation}
where ${\rm FWHM}$ is the full-width at half-maximum of the image point-spread function (PSF) and $Q_{\gamma}$ is the signal-to-noise ratio with which the star is detected.  Assuming a diffraction limited, Nyquist-sampled PSF and a circular unobscured aperture, FWHM $\simeq 1.22 (\lambda/D)$ where $\lambda$ is the wavelength and $D$ is the diameter.  For source-limited photon noise\footnote{The contribution of the background flux and detector noise to the astrometric uncertainty is generally much smaller than that due to the source, even for the faintest viable reference stars}, $Q_\gamma = \sqrt{N_\gamma}$, where $N_\gamma=\epsilon \Gamma_\gamma \pi (D/2)^2 \Delta \lambda t_{\rm exp}$ is the number of photons collected for the source, $\epsilon$ is the overall photon collection efficiency, $\Gamma_\gamma$ is the photon collection rate (per unit area per unit time per unit wavelength), $\Delta \lambda$ is the filter width and $t_{\rm exp}$ is the exposure time.  

In general, the target star will have a magnitude of $V\sim 4-6$, whereas the reference stars will primarily be fainter than $V\sim 15$, and thus be $\gtrsim 10^4$ times fainter than the target star.  We therefore expect the astrometric uncertainty to be dominated by the photon noise due to the total photon noise from the reference stars, even for a relatively large number of reference stars. We estimate the astrometric precision to be
\begin{equation}
    \begin{split}
      \sigma_{\rm ast} \simeq 1.8~\mu{\rm as}~
10^{-0.2(V-15)}
\left(\frac{t_{\rm exp}}{30~{\rm min}}\right)^{-1/2}
\left(\frac{D}{6~{\rm m}}\right)^{-1}\\ \quad
        \left(\frac{\lambda/D}{20~{\rm mas}}\right)
\left(\frac{\epsilon}{0.25}\right)^{-1/2}.
    \end{split}
\end{equation}
It should be noted that this scales as $D^{-2}$. There is one power of $D$ from the angular size of the PSF and one power of $D$ from the collecting area. 

If one has $N_{\rm ref}$ stars in the field of view, each with a given $V$ magnitude, then the astrometric uncertainty per epoch is
\begin{equation}
    \sigma_{\rm ref} = \frac{\sigma_{\rm ast}(V)}{N_{\rm ref}^{1/2}}
\end{equation}
Thus, $\sim 36$ reference stars with $V=15$ are needed to achieve astrometric precision per epoch of $\sim 0.3~\mu{\rm as}$ with this exposure time.  

Assuming a circular, face-on orbit, and $N_{\rm obs}$ measurements uniformly spaced over a time $T$ with $P\ll T$, the precision with which the mass of the planet can be measured is
\begin{equation}
    \frac{\sigma_{M_p}}{M_p} \simeq 10\% 
    \left(\frac{N_{\rm obs}}{100}\right)^{-1/2}
    \left(\frac{\sigma_{\rm ref}}{{0.3}~\mu{\rm as}}\right)
    \left(\frac{M_p}{M_\oplus}\right)^{-1}
     \left(\frac{M_*}{M_\odot}\right).
\end{equation}
The number of reference stars with a magnitude within $\Delta V/2$ of $V$ in the field of view is
\begin{equation}
N_{\rm ref} \simeq \Sigma_*(V) \Omega_{\rm det} \Delta V,
\end{equation}
where $\Sigma_*(V)$ is the angular surface density of stars (number of stars per steradian per magnitude) and $\Omega_{\rm det}$ is the solid angle on the sky covered by the detector.   Note that when $\Sigma_*(V) \Omega_{\rm det} \Delta V \sim 1$, Poisson fluctuations in the number of reference stars that are located in the field of view will become important.  

The angular surface density of stars depends on where in the sky the target star is located.  The target stars for the HWO survey of potentially habitable planets orbiting nearby stars will be nearly uniformly spread over the sky because their typical distances are $\lesssim 20~{\rm pc}$, which is significantly less than the scale height of even early-type stars of $\sim 100~{\rm pc}$.


\begin{figure*}[ht!]
   \centering
       \includegraphics[width=0.49\linewidth]{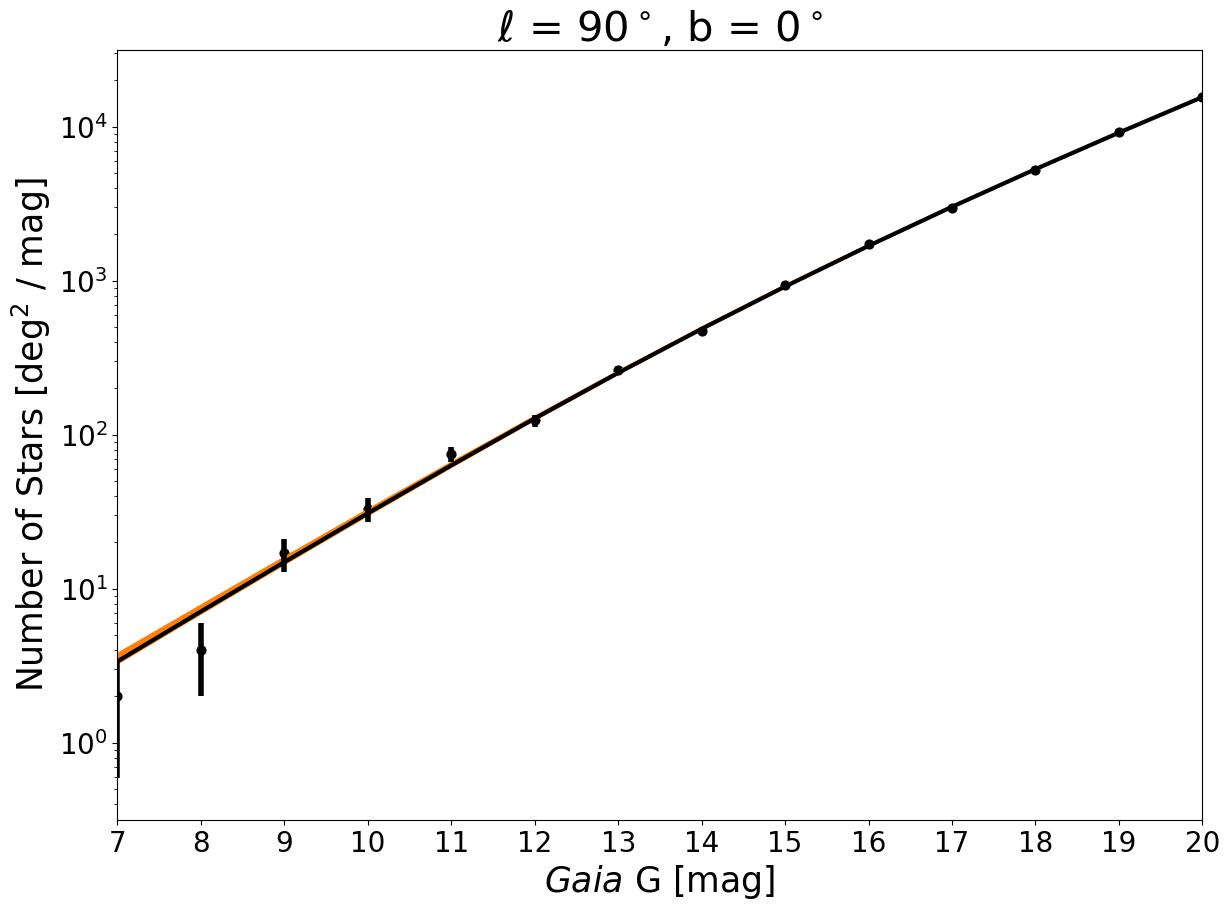}
        \includegraphics[width=0.49\linewidth]{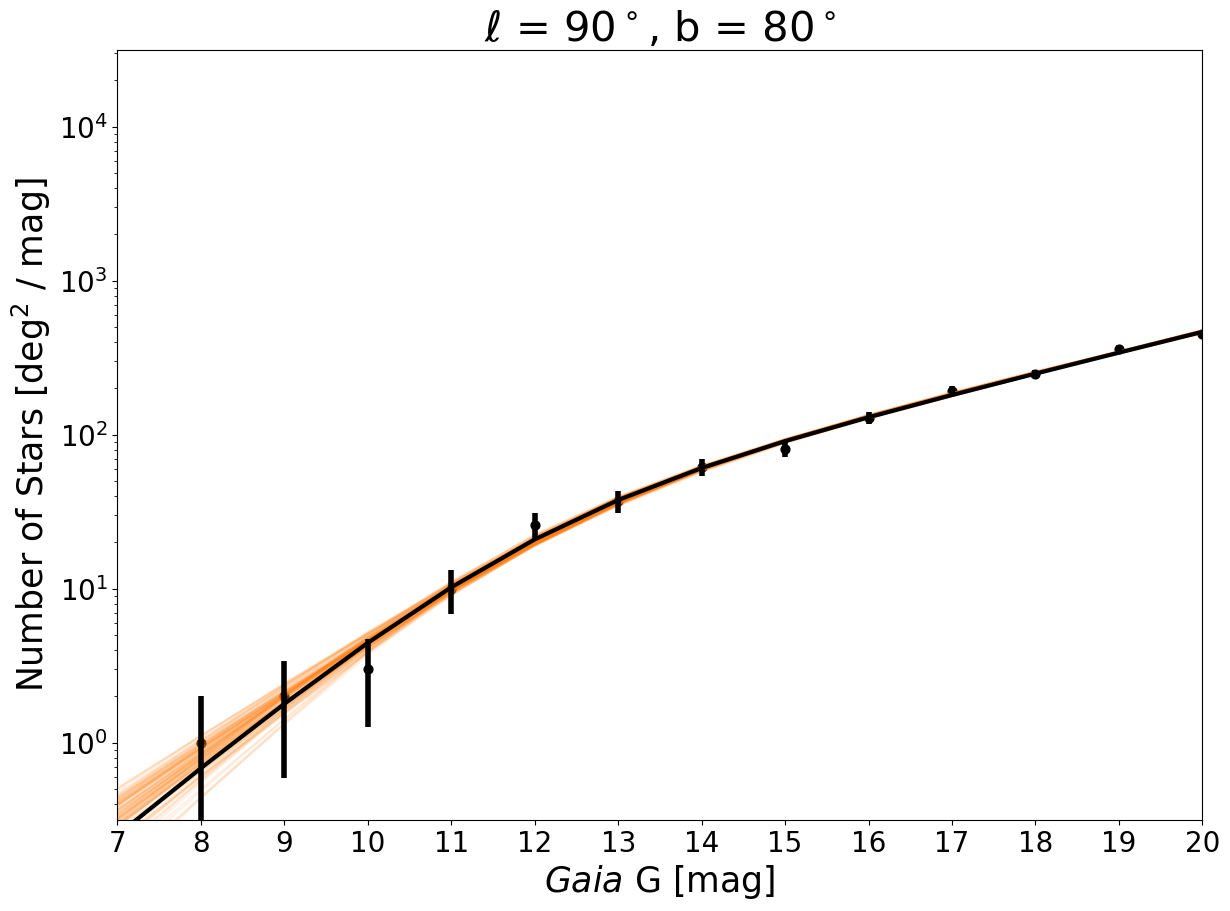}
        \caption{\small {\bf Number of stars per square degree per magntitude in the Gaia $G$ band} as function of $G$ magnitude for a line-of-sight toward Galactic latitude $l=90^\circ$ and Galactic longitude $b=0^\circ$ (left), and $b=80^\circ$ (right)
 }\label{fig:refstarsG}
    \end{figure*}
    
Figure \ref{fig:refstarsG} shows the number of stars per square degree per $G$ magnitude as a function of magnitude for a line of sight toward Galactic latitude $l=90^\circ$ and for two different Galactic longitudes, one in the plane ($b=0^\circ$) and one near the Galactic poles ($b=80^\circ$).  These were calculated using \texttt{SYNTHPOP} Galactic model \citep{Kluter:2025}, which reproduces the actual Gaia number counts to within a factor of $\sim 2$.  We adopted this model rather than using the actual Gaia number counts in order to easily explore the trade of using different filters for astrometric measurements, as \texttt{SYNTHPOP} can be used to predict the stellar surface density in most standard filters \citep{Kluter:2025}.  

In the plane, there are a sufficient number of reference stars.  There are roughly $\sim 10^3$ stars per square degree within a bin of $\Delta G=1$ at $G\simeq 15$, and so $\sim 10$ in a FOV of 6'$\times$6'.  Of course, brighter and fainter stars contribute to reducing the astrometric uncertainty as well. Near the pole, the surface density of reference stars is much lower\footnote{In particular, there is a break in the magnitude distribution of stars for lines-of-sight away from the Galactic plane because of the finite scale height of the disk.  Magnitude-limited number counts of stars are generally dominated by FGK-type main sequence stars in the range $G=10-16$.  Taking a typical $G$-band absolute magnitude of $M_G\simeq 5$, we expect a break in the magnitude distribution at $G_{\rm br}\simeq M_G + 5\log(H_z/\sin{b})-5$ where $H_z\simeq 300~{\rm pc}$ is the vertical scale height of thin-disk FGK main sequence stars, and thus $G_{\rm br}\sim 12.5$, roughly consistent with the location of the break for $b=80^\circ$ in the right panel of Figure \ref{fig:refstarsG}}: there are only $\sim 10^2$ stars per square degree within a bin of $\Delta G=1~{\rm mag}$ at $G\simeq 15$, and thus only $\sim 1$ in a FOV of 6'$\times$6'.  Although using fainter reference stars will improve the astrometric uncertainty, as we will show, they are on average not sufficiently plentiful to yield a value of $\sigma_{\rm ref}\sim 0.3~\mu{\rm as}$ for a $30~{\rm min}$ exposure on a 6m aperture HWO.


 \begin{figure*}[ht!]
 \centering
       \includegraphics[width=0.48\linewidth]{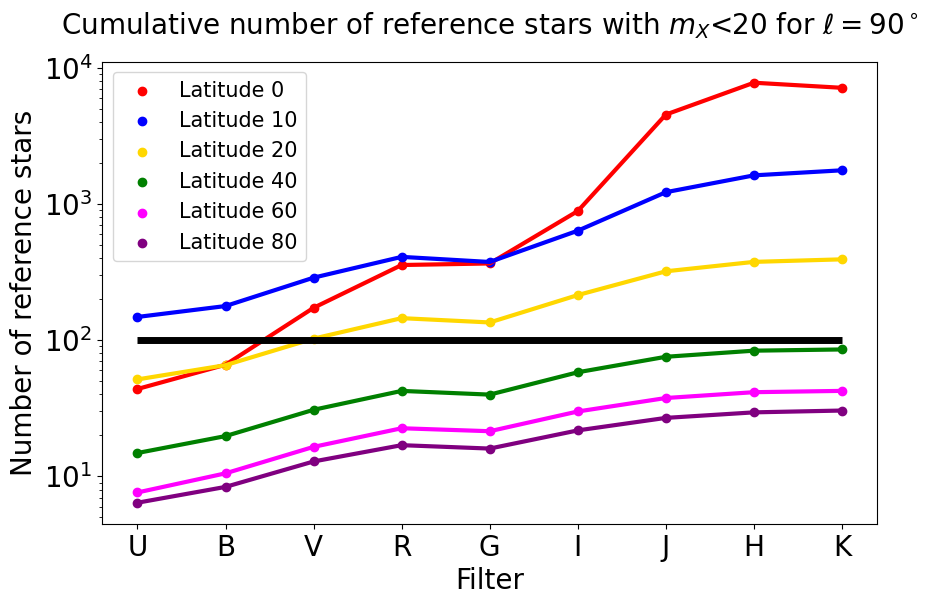}
        \includegraphics[width=0.48\linewidth]{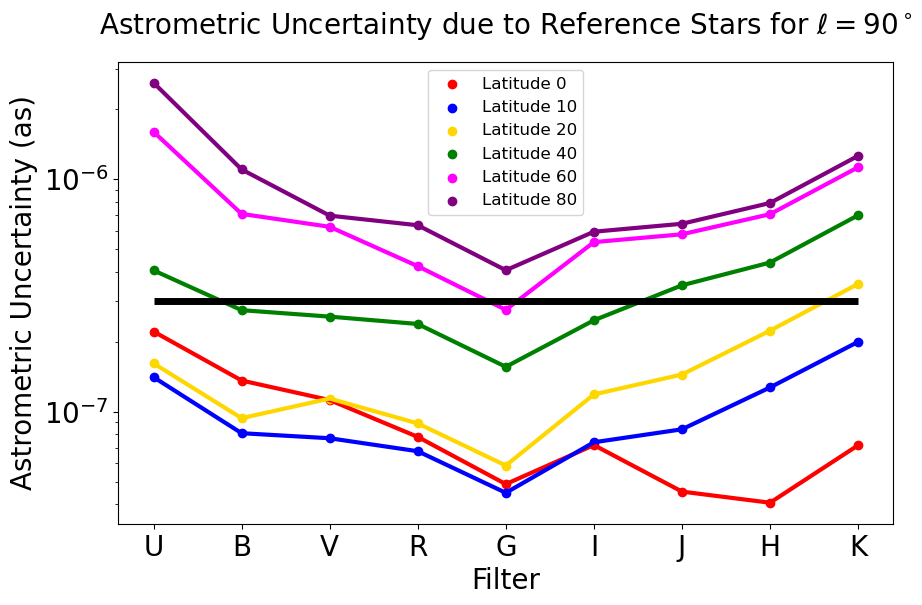}
        \caption{\small (left) Cumulative number of reference stars with magnitude $m_X\le20$ as a function of filter for $\ell = 90^\circ$ and various Galactic latitudes $b$ for a detector FOV of $\Omega=36~{\rm arcmin^2}$.  (right) The contribution of the astrometric uncertainty due to reference stars as a function of filters for $\ell = 90^\circ$ and various Galactic latitudes $b$.  Here we have assumed a diffraction-limited PSF, $t_{\rm exp}=30~{\rm min}$, $D=6~{\rm m}$, $\Omega_{\rm det}=36~{\rm arcmin^2}$, and an overall throughput of $\epsilon=0.25$.
 }\label{fig:refstars}
    \end{figure*}

We also explore the trade-off of the filter used to take the astrometric measurements.  We consider the standard Johnson-Cousins UBVRIJHK filters as well as the wider Gaia G filter. Here, there are two competing effects.  For shorter wavelengths, the width of a diffraction-limited PSF decreases as $\lambda$, and thus the astrometric uncertainty decreases by the same amount for a fixed signal-to-noise ratio (see Table~\ref{tab:filters}).  However, the number of reference stars down to a given magnitude generally increases for shorter wavelengths, as shown in Figure \ref{fig:refstars}, although the details of this depend on Galactic latitude because of the competing effects of the different scale heights of stars of different temperatures and wavelength-dependent extinction. 

We estimate the total astrometric uncertainty due to all reference stars as
\begin{equation}
    \sigma_{\rm ref} = \left[ \sum_i^{N_{\rm mag}} \frac{N_{\rm ref}(m_{X,i})}{\sigma_{\rm ast}^2(m_{X,i})} \right]^{-1/2}
\end{equation}
where the sum is over the $N_{\rm mag}$ magnitude bins with $m_{X,i}=7,8,9,10,..., 20$, the number of reference stars per magnitude bin is $N_{\rm ref}(m_{X,i})=\Sigma_*(m_{X,i})\Omega_{\rm det} \Delta m_{X}$ and $\Delta m_X=1~{\rm mag}$.  We conservatively exclude bins where the expected number of reference stars is $N_{\rm ref}\le 1$.  This astrometric uncertainty is shown in Figure~\ref{fig:refstars} for $t_{\rm exp}=30~{\rm min}$, $D=6~{\rm m}$, $\Omega_{\rm det}=36~{\rm arcmin^2}$, and an overall throughput of $\epsilon=0.25$.  Perhaps not surprisingly, optimal filters are centered around V, R, and G ($\lambda_{\rm eff} \sim 500-700~{\rm nm}$), where the competing effects of increasing the astrometric uncertainty per star with wavelength and the increase in the number of reference stars at a given magnitude with wavelength result in a broad minimum in astrometric uncertainty as a function of wavelength. This is essentially independent of location in the sky except for very close to the Galactic plane, where the lower extinction in the near-IR results in a significant `bump' in the number of reference stars. 

We find that the optimal filter is the Gaia G band.  This is because of the wavelength trade-off discussed above and the fact that it is a relatively wide filter, with $\Delta \lambda_{\rm FWMH}/\lambda_{\rm eff} \sim 0.65$ compared to $\sim 0.15-0.2$ for V and R.  However, it is worth noting that a broad filter may pose other challenges, and the target star is likely to be significantly bluer than the reference stars, and thus color-dependent distortions may be more significant. 

For the G band and our fiducial assumptions ($t_{\rm exp}=30~{\rm min}$, $D=6~{\rm m}$, $\Omega_{\rm det}=36~{\rm arcmin^2}$, $\epsilon=0.25$), the contribution to the astrometric uncertainty per epoch due to the reference stars is $\lesssim 0.3~\mu{\rm as}$ for essentially all lines of sight except for very close to the Galactic poles.  

\subsection{Total Observing Time and Observation Strategy Considerations}\label{sec:totobs}

Finally, in order to achieve a mass measurement precision of $\sim 10\%$, roughly $N_{\rm abs}=100$ measurements, each with an astrometric uncertainty of $\sigma_{\rm ref}=0.3~\mu{\rm as}$ are needed.  This leads to a total observation time of
\begin{equation}
    \begin{split}
        T_{\rm tot}= N_{\rm tar} \eta_{\oplus} N_{\rm obs} (t_{\rm exp}+t_{\rm over}) 
    \simeq 170~{\rm days} 
    \left(\frac{ N_{\rm tar}}{165}\right)\\ \quad
        \left(\frac{\eta_{\oplus}}{0.25}\right)
      \left(\frac{N_{\rm obs}}{100}\right)
       \left(\frac{t_{\rm exp}+t_{\rm over}}{1~{\rm hr}}\right),
    \end{split}
\end{equation}
where $N_{\rm tar}=164$ is the number of stars in the HWO Tier 1 target list in \citealt{Tuchow2025} (which was adapted from \citealt{mamajek:2024}), and we have assumed an overhead time of $30~{\rm min}$.  

Note that in Figure \ref{fig:refstars}, the lines of sight with $b\lesssim 40^\circ$ will have a sufficient number of reference stars so that the astrometric precision of the photon noise will be significantly lower than $\sim 0.3~\muas$ for an exposure with $t_{\rm exp} = 30~{\rm min}$.  Therefore, one might imagine an observation strategy in which one acquires fewer than $100$ epochs to achieve a 10\% mass measurement.  In reality, the strategy is constrained by the fact that there will be a systematic floor in the astrometric uncertainty that can be achieved per epoch that is likely to be of the order or even greater than $\sim 0.3~\muas$.   On the other hand, one could imagine exposing for a shorter amount of time per epoch with the same number of epochs to decrease the total observing time.  However, the total amount of mission time will ultimately be limited by overheads, in the limit of short exposure times.   Finally, larger apertures will result in improved astrometric precisions per epoch for a fixed exposure time, but these gains can only be realized if the systemics can be controlled to the same level.    

A detailed study of these observation strategy trades that account for realistic overhead times, systematics, and the actual set of reference stars for each target, is needed.  

\begin{table*}[ht!]
    \centering
    \begin{tabular}{|c|cc|c|cc|c|c|c|}
    \hline
       Filter & $\lambda_{\rm eff}$  & $\Delta\lambda_{\rm FWHM}$ & $\theta_{\rm diff}$ & SNR & $\sigma_\mathrm{ast}$  & Pixels & $\Omega_{\rm det}$ & $0.3~\mu as/\theta_{\rm pix}$\\
            \ & [$\mu$m]             &  [$\mu$m]                  & [mas]             &     & [$\mu$as]              & [Mpix] & ($N_{\rm pix}$=1~Gpix)                          & [$\times 10^4$]\\
              &      &               &                            & \multicolumn{2}{c|}{$m_X$=15 (Vega)}             & 36 arcmin$^2$ & [arcmin$^2$] & \\
       \hline
U & 0.360 & 0.060 & 12.4 & 7600 & 1.63 & 3,384 & 10.6 & 4.8 \\
B & 0.438 & 0.090 & 15.1 & 12600 & 1.19 & 2,286 & 15.7 & 4.0 \\
V & 0.545 & 0.085 & 18.7 & 10400 & 1.81 & 1,476 & 24.4 & 3.2 \\
R & 0.641 & 0.150 & 22.0 & 11600 & 1.90 & 1,067 & 33.7 & 2.7 \\
G & 0.639 & 0.455 & 22.0 & 18200 & 1.21 & 1,074 & 33.5 & 2.7 \\
I & 0.798 & 0.150 & 27.4 & 9300 & 2.95 & 688 & 52.3 & 2.2 \\
J & 1.220 & 0.260 & 41.9 & 8000 & 5.25 & 294 & 122.2 & 1.4 \\
H & 1.630 & 0.290 & 56.0 & 5900 & 9.56 & 165 & 218.1 & 1.1 \\
K & 1.190 & 0.410 & 40.9 & 4800 & 8.58 & 309 & 116.2 & 1.5 \\
    \hline
    \end{tabular}
    \caption{\small Column 1 gives the filter bandpass, columns 2 and 3 give the effective wavelength and full-width at half-maximum of the filter. Columns 3-5 give the approximate diffraction limit, signal-to-noise ratio, and single-epoch photon noise limited astrometric precision for a $D=6~{\rm m}$ telescope, a total exposure time per epoch of $t_{\rm exp}=30~{\rm min}$, and overall system throughput of $\epsilon=0.25$, for various standard filters.  Column 5 gives the approximate number of pixels required assuming Nyquist sampling of the PSF ($=\lambda/D/2$) and a 6'$\times$6' field-of-view.  Column 6 gives the FOV assuming a Nyquist sampled camera with 1 Gigapixels.  Column 7 gives the expected astrometric signal of an Earth analog around a solar-type star in units of the angular size of the pixel $\theta_{\rm pix}$.}
    \label{tab:filters}
\end{table*}

\subsection{Basic Engineering Constraints}\label{sec:goals}

Here we briefly discuss some of the basic engineering considerations for achieving the photon noise accuracy estimated above.  To achieve the highest astrometric accuracy, the imaging camera should be diffraction limited at the wavelengths of the observations. Furthermore, Nyquist sampling of the PSF is preferred, as undersampling the PSF will degrade the resolution. For a field of view of $36$~acrmin$^2$, to achieve this requires a camera with a pixel size of $\theta_{\rm pix}\sim \lambda/D/2 \sim 11~{\rm mas}$ and $N_{\rm pix}\sim 1$~Gpix in the G band.  To achieve a precision $\sim 0.3\muas$ per epoch, systematic errors must be controlled at a fraction of $0.3~\muas/\theta_{\rm pix} \sim 3\times 10^{-4}$ of a pixel in the G band.  

There are competing effects on the camera requirements for the other wavelengths.  The required number of pixels decreases for longer wavelengths, but the level at which the systematics must be controlled becomes a larger fraction of a pixel for shorter wavelengths.  In addition, achieving diffraction-limited imaging becomes increasingly more difficult at shorter wavelengths.  

A key error term in the astrometric error budget is the optical distortion which change from epoch to epoch as the observatory deforms with thermal gradients, composites desorption, and radiation compactation. Optical distortion bias the position of the background reference stars causing an error in the astrometric measurement. Preliminary calculations shows that the telescope, thanks to the laser metrology, will be stable enough to provide astrometric accuracy in the order of $1\mu$as, however, the telescope laser metrology should be able to maintain the stability over years time scale to provide proper orbital sampling. The intermediate optics and the HRI are likely to contribute a larger term to the astrometry error budget. A calibration scheme such as \citep{Guyon:2012, Bendek:2020} could be used to reduce this error term.


\bibliography{author.bib}

\end{document}